# A Comparison of the Cc and R3c Space Groups for the Superlattice Phase of Pb(Zr$_{0.52}$Ti$_{0.48}$)O$_3$

Ragini, Akhilesh Kumar Singh, Rajeev Ranjan and Dhananjai Pandey*

School of Materials Science and Technology, Institute of Technology,
Banaras Hindu University, Varanasi-221005, India.

*email: dpandey@bhu.ac.in

## Abstract

Recent controversy about the space group of the low temperature superlattice phase of Pb(Zr$_{0.52}$Ti$_{0.48}$)O$_3$ is settled. It is shown that the R3c space group for the superlattice phase cannot correctly account for the peak positions of the superlattice reflections present in the neutron diffraction patterns. The correct space group is reconfirmed to be Cc. A comparison of the atomic coordinates of Cc and Cm space groups is also presented to show that in the absence of superlattice reflections, as is the case with XRD data, one would land up in the Cm space group. This superlattice phase is found to coexist with another monoclinic phase of Cm space group.

## I. INTRODUCTION:

Ever since its discovery, the perovskite lead zirconate titanate Pb(Zr$_x$Ti$_{1-x}$)O$_3$ (PZT) has remained by far the most widely used piezoelectric ceramic material for transducer and actuator applications [1]. Recently, Noheda et al [2-3] showed that the tetragonal compositions with x=0.50, 0.51 and 0.52 transform to a monoclinic phase with Cm space group. In a subsequent comprehensive study of the temperature dependence of the piezoelectric resonance frequency, dielectric constant and crystal structure using x-ray, electron and neutron diffraction data, Ragini et al [4] and Ranjan et al [5] discovered yet another low temperature monoclinic phase, which is a superlattice of the Cm phase discovered earlier by Noheda et al [2]. It was also shown that this superlattice phase results from an antiferrodistortive phase transition involving antiphase tilting of oxygen octahedra in the Cm phase [5] and its space group is Cc [6]. Recently, Frantti et al [7], while confirming the reports by Ranjan et al [5] about the appearance of a superlattice phase of PZT using powder neutron diffraction data, have proposed R3c space group for this phase in analogy with the R3m to R3c transition [8] on the Zr-rich side of the morphotropic phase boundary (MPB) in PZT. These workers have also proposed that this R3c superlattice phase coexists with the monoclinic Cm phase discovered earlier by Noheda et al [2]. On the otherhand, Noheda et al [9] carried out a TEM study of the low temperature transitions in PZT with x=0.520 and confirmed the findings of Ragini et al [4] about the appearance of a superlattice phase and proposed the same space group (Cc), which was earlier established by Hatch et al [6]. However, according to Noheda et al [9], the superlattice phase with Cc space group is a minority monoclinic phase, which coexists with the majority monoclinic phase with Cm space group discovered earlier [2]. Noheda et al [9] have proposed that the superlattice phase of PZT results from local internal inhomogeneities or local stresses. There is thus a need to settle the existing controversy about the origin of the superlattice reflections observed in PZT with $x = 0.520$ at low temperatures. In particular, it needs to be settled whether the superlattice reflections are due to a monoclinic phase with Cc space group as proposed by Hatch et al [6], or due to a rhombohedral phase with R3c space group as proposed by Frantti et al [7]. The second issue to be settled is whether or not there is another coexisting phase as proposed by Frantti et al [7] and Noheda et al [9]. The third issue to be settled is that if there is a coexisting Cm phase with the superlattice phase, is it pseudo-rhombohedral type or pseudo-tetragonal type. Using Rietveld analysis of powder x-ray diffraction data, Ragini et al [10] have reported that the monoclinic Cm phase which coexists with the tetragonal P4mm phase at room temperatures is of pseudo-rhombohedral type whereas Noheda et al [2] have reported a pseudo-tetragonal type monoclinic Cm phase at low temperatures.

In order to settle these issues, we have reexamined the low temperature structure of PZT with x=0.520 using Rietveld analysis of neutron powder diffraction data keeping in mind the above possibilities as structural solutions. We have used the same powder neutron diffraction data that was used earlier [5, 6]. It is shown that a pseudo-tetragonal superlattice Cc phase coexisting with a pseudo-rhombohedral Cm phase accounts most satisfactorily for the observed features in the powder neutron diffraction data. The R3c space group proposed by Frantti et al [7] is shown to give incorrect peak positions for the superlattice reflections. Using appropriate coordinate transformations, it is also shown that the refined parameters of the Cm phase obtained earlier by Noheda et al [2] by using XRD data at 20K are in agreement with those of the Cc phase obtained by us, except for the sense of displacements along the monoclinic [010] which is responsible for the intensity of the superlattice peaks present in the powder neutron diffraction data. This sense of displacements along the monoclinic [010] could not be captured in the XRD work of Noheda et al [2] since the superlattice peaks are not discernible in the XRD data.

## II. REFINEMENT DETAILS:

Structure refinement was performed using Rietveld refinement programme DBWS 9411 [11]. The background was estimated by linear interpolation between fixed values. A pseudo-Voigt function was chosen to generate profile shape for the neutron diffraction peaks. The Cc space group has only one Wyckoff site symmetry 4(a) with general asymmetric unit of the structure consisting of 5 atoms of which we fix, by convention, Pb at (0.00, 0.25, 0.00). The remaining atoms have coordinates as follows: Zr/Ti at $(0.25+\delta x_{Ti}, 0.25+\delta y_{Ti}, 0.75+\delta z_{Ti})$ and three oxygen atoms, O1 at $(0.00+\delta x_{O1}, 0.25+\delta y_{O1}, 0.50+\delta z_{O1})$, O2 at $(0.25+\delta x_{O2}, 0.50+\delta y_{O2},$ $0.00+\delta z_{O2}$) and O3 at ($0.25+\delta x_{O3}$, $0.00+\delta y_{O3}$, $0.00+\delta z_{O3}$). The various δ's represent the refinable parameters. There are four atoms in the asymmetric unit of the monoclinic phase with Cm space group. Pb is fixed at (0.00, 0.00, 0.00), Zr/Ti at ($0.50+\delta x_{Ti/Zr}$, 0.00, $0.50+\delta z_{Ti/Zr}$), O1 at ($0.50+\delta x_{O1}$, 0.00, $0.00+\delta z_{O1}$) and O2 at ($0.25+\delta x_{O2}$, $0.25+\delta y_{O2}$, $0.50+\delta z_{O2}$). Following Megaw and Darlington [12], the asymmetric unit of rhombohedral (R3c) structure with hexagonal axes consists of Pb at (0.00, 0.00, $0.25+\delta z_{Pb}$), Zr/Ti at (0.00, 0.00, $0.00+\delta z_{Zr/Ti}$) and O at ($1/6+\delta x_O$, $1/3+\delta y_O$, $1/12$).

### III. RESULTS AND DISCUSSION:

**A. Selection of the correct structural model for the superlattice phase:**

To make a choice among different structural models for the superlattice phase, the fits between observed and calculated profiles, as obtained by the Rietveld technique, were compared for different models. In the earlier refinement by Hatch et al [6] considering pure Cc space group, the following constraints were used, $\delta y_{Ti/Zr} = \delta y_{OI} = 0$, $\delta x_{OII} = \delta x_{OIII}$, $\delta y_{OII} = -\delta y_{OIII}$ and $\delta z_{OII} = \delta z_{OIII}$. In the present work, we realized that even without imposing these constraints, refinement converged smoothly with nearly identical parameters. Fig.1 shows the fits between observed and calculated profiles for the 200 and 222 pseudocubic elementary perovskite reflections, and 311 and 511 superlattice reflections, of PZT with x=0.520 at 10K obtained after Rietveld refinements using the following structural models: (i) pure Cc phase (ii) coexistence of Cm and R3c phases (iii) coexistence of Cc and Cm phases (model I) (iv) coexistence of Cc and Cm phases (model II). The main difference between the two Cc+Cm coexistence models is that in model II, the Cc phase has a pseudo-tetragonal character while Cm has a pseudo-rhombohedral character while it is otherway round for model I. As can be seen from Fig.1, pure Cc space group model accounts satisfactorily for the 311 and 511 superlattice reflections (with respect to the doubled pseudocubic cell $2a_p \times 2b_p \times 2c_p$) abut considerable mismatch between the observed and calculated profiles is seen for the pseudocubic 200 and 222 profiles. Evidently, there is sufficient scope of improvement in the fit between the observed and calculated intensities for the 200 and 222 profiles and a two-phase structural model may be desirable. The Cm +R3c phase coexistence model, proposed by Frantti et al [7], gives better fit for the 200 and 222 pseudocubic profiles in comparison to that of pure Cc phase but distinct mismatch between the peak positions of the observed and calculated profiles for the 311 and 511 superlattice reflections is observed (see Fig.1 for Cm+R3c coexistence model). The calculated peak positions of the superlattice reflections occur at lower 2θ angles as compared to the observed ones. The most marked difference can be seen at the 511 superlattice peak; the observed peak position of this superlattice reflection occurs at 2θ = 64.60° whereas the calculated peak occurs at 2θ = 64.02°, a difference of 0.6° which can be easily noticed even on a medium resolution powder neutron diffractometer. A careful inspection of the inset in Fig. 3 of Frantti et al [7] also clearly reveals identical mismatch between the observed and calculated peak positions using R3c space group. Attempt to force the match in the peak position of the superlattice reflections for the R3c space group resulted in shifting of main perovskite reflections corresponding to the R3c phase towards higher two-theta side and consequently large mismatch between the observed and calculated profiles for the perovskite reflections. The peak position of the 511 superlattice reflection could be matched only when the 200 peak of the R3c phase reaches the position of the 220 peak of Cm phase in this model, but it leads to very large difference between observed and calculated profiles for the main perovskite peaks. Thus the (Cm+R3c) phase coexistence model for the structure of PZT with x=0.520 at 10K is not acceptable. This at least settles that the choice of the rhombohedral cell is incorrect for indexing the superlattice reflections appearing in the neutron powder diffraction pattern of the low temperature phase of PZT with x=0.520. The fitted pattern with Cc space group by Hatch et al [6] did not have such a mismatch between the observed and calculated positions of the superlattice reflections, implying that the unit cell corresponding to the Cc space group is the correct cell for exact indexing of the weak superlattice reflections in the powder neutron diffraction pattern. In the light of this subtle but very important fact, the doubt raised by Frantti et al [7] about the correctness of the Cc space group no longer holds valid.

After ruling out the Cm+R3c phase coexistence model, we considered the coexistence of the monoclinic Cm and monoclinic Cc phases. It is evident from Fig.1 that the fit between observed and calculated profiles is identical for the models I and II of the phase coexistence of the Cm and Cc phases, for the 200 and 222 pseudocubic elementary perovskite reflections. For model I, when Cc phase is considered to be pseudo-rhombohedral, which in turn requires the coexisting Cm phase to be pseudotetragonal to account for the splitting of 200 type perovskite peaks, the calculated superlattice peak positions are seen to occur at lower $2\theta$ angles as compared to the observed ones, similar to the Cm+R3c phase coexistence model. However, for model II, where the Cc phase is considered to be pseudotetragonal (similar to that reported by Hatch et al [6]), very good match between observed and calculated profiles is obtained for both the main perovskite reflections as well as the superlattice peaks. It is also clear that the consideration of pseudotetragonal or pseudorhombohedral type Cc phases does not make any difference as far as the fits for the main perovskite reflections are concerned. However, the match for the peak positions of the superlattice reflections are quite sensitive to the type of the Cc phases.

Thus, we can conclude that PZT with x=0.520 at 10K consists of a mixture of two phases with Cc and Cm space groups and not R3c and Cm space groups proposed by Frantti et al [7]. The elementary perovskite subcell parameters $a_p=c_m/\sqrt{2} = 4.044$Å, $b_p=b_m/\sqrt{2} = 4.032$Å, $c_p=1/2 (a_m^2 + c_m^2 - 2a_m c_m \cos(\pi-\beta))^{1/2} = 4.129$ Å of the Cc phase show pseudotetragonal character ($a_p \approx b_p \neq c_p$) while those $a_p=a_m/\sqrt{2}=4.072$ Å, $b_p=b_m/\sqrt{2}=4.051$ Å, $c_p=c_m=4.093$ Å, $\beta \approx 90°$ of the coexisting Cm phase are pseudorhombohedral ($a_p{\approx}b_p{\approx}c_p$). This coexisting Cm phase is similar to that reported by Ragini et al [10] in PZT (x=0.520) at room temperature which coexists with the tetragonal phase. Fig. 2 depicts the observed, calculated and the difference plots for Cc+Cm phase coexistence (model II) in the two-theta range 20 to 120 degrees. As can be seen from this figure very good fit between the observed and calculated profiles is obtained. The refined structural parameters for the Cc and Cm phases obtained after the Rietveld refinement is given in Table 1. The atomic coordinates of the Cc phase given in Table 1 and those reported by Hatch et al [6] are comparable. Further, refinement clearly reveals that the Cc phase is a majority phase (65%) while Cm is a minority phase (35%). Thus the observation of Noheda et al [9], based on their TEM study of thin regions, that Cc phase is a minority phase, is not supported by the analysis of the bulk neutron powder diffraction data.

**B. Comparison of the refined structural parameters of the Cc phase with those obtained by Noheda et al [2] using Cm space group.**

It is worth comparing the refined coordinates of the Cc phase given in Table 1 with those obtained by Noheda et al [2] for x = 0.520 at 20K using X-ray synchrotron data assuming Cm space group. As pointed out earlier by Ragini et al [4] and Ranjan et al [5], the superlattice reflections of PZT at this temperature are discernible on electron and neutron diffraction patterns only and not on the powder XRD patterns. In the absence of the superlattice peaks in the x-ray synchrotron data, the most plausible space group which can account for the entire powder diffraction patterns was shown to be Cm by Noheda et al [2]. The unit cell axes of the Cm ($a_m$, $b_m$, $c_m$,) and the Cc ($a_c$, $b_c$, $c_c$,) space groups are related through the following transformation matrix (M):

$$\begin{pmatrix} a_m \\ b_m \\ c_m \end{pmatrix} = (M) \begin{pmatrix} a_c \\ b_c \\ c_c \end{pmatrix}, \text{ where}$$

$$(M) = \begin{pmatrix} 0 & 0 & -1 \\ 0 & 1 & 0 \\ 1/2 & 0 & 1/2 \end{pmatrix} \quad \ldots\ldots\ldots(1)$$

The atomic coordinates ($x_c$, $y_c$, $z_c$) given in Table 1 for the Cc space group can therefore be transformed to ($x_m$, $y_m$, $z_m$) corresponding to the Cm space group via the following transformation equation:

$$\begin{pmatrix} x_m \\ y_m \\ z_m \end{pmatrix} = \left((M)^{-1}\right)^T \begin{pmatrix} x_c \\ y_c \\ z_c \end{pmatrix}, \text{ where} \quad \ldots\ldots\ldots (2)$$

$$\left((M)^{-1}\right)^T = \begin{pmatrix} 1 & 0 & -1 \\ 0 & 1 & 0 \\ 2 & 0 & 0 \end{pmatrix} \quad \ldots\ldots\ldots (3)$$

This gives:

$$x_m = x_c - z_c, \quad y_m = y_c, \quad z_m = 2x_c \quad \ldots\ldots\ldots (4)$$

Since Pb is kept at origin in the Noheda et al's [2] model, we need to shift the $y_m$ coordinate through $-b/4$. The equivalent $x_m$, $y_m$, $z_m$ coordinates of the 5 atoms in the asymmetric unit of the Cc space group listed in Table 1 are given in Table 2 after incorporating this shift of origin. The same table also lists the coordinates obtained by Noheda et al [2] using the XRD data.

The asymmetric unit of the Cc space group has one additional O atom as compared to the asymmetric unit of the Cm space group. This is due to the fact that O2 and O3 atoms are related through a mirror at y=1/2 in the Cm space group but not in the Cc space group. As a result, the coordinates of O3 for the Cc space group given in Table 2 should be compared with the

9coordinates of the equivalent atom O2 in the Cm space group after adding (0, 0.5132, 0). A comparison can now be made between the coordinates for the Cc (proposed by us) and Cm (proposed by Noheda et al [2]) space groups given in Table 2. It follows that the refined x and z coordinates of all the atoms for the Cc space group are very close to those obtained by Noheda et al [2]. This shows that the structure of the low temperature monoclinic ($F_M^{LT}$) phase (Cc space group) retains the main structural framework of the higher temperature monoclinic ($F_M^{HT}$) phase with Cm space group. The only difference is in terms of the sense of displacements in the y-direction of all the atoms as explained below.

For easy comparison with the unit cell given by Noheda et al [2], the Cc space group can equivalently be written as Ia in a non-standard setting. The unit cell of the Ia setting contains two subcells of Cm type stacked along [001] [13]. The coordinates given in Table 2 actually correspond to one of the subcells of Cm type. We can now compare the sense of displacements with respect to the symmetric ideal perovskite positions of various atoms in the Ia and Cm space groups with the help of Fig. 3. The arrows in this figure show the sense of displacements (and not magnitudes) of various atoms in each layer along the unit cell axes contained in the layers under consideration. For Ti/Zr the magnitude of the y-displacement is within the e.s.d (see Table 2) and hence ignored in Fig. 3a. The Ia unit cell contains four layers of atoms at z=0, 1/4 , 1/2 and 3/4 as shown in Fig. 3(a) while the Cm unit contains two layers at z=0 and z=1/2. The z=1/2 layer of Cm is to be compared with z=1/4 layer of Ia since the 'z' coordinates listed in Table 2 are doubled for easy comparison with the true Cm phase coordinates. It is evident from Fig. 3(a) that the sense of atomic displacements of O and Ti atoms within a layer are related through 'a'-glide at y=1/4, 3/4; e.g. the pair of O atoms 2 –2′ and 3 - 3′ in Fig. 3a are a-glide related. The O and Ti atoms in the Cm unit cell are related through mirror at y= 1/2; e.g. see O atoms labeled 2 and 3 in Fig. 3b. When viewed along the elementary perovskite pseudocubic [100], [010], and [001] directions, it is evident that the y-displacements of neighbouring O atoms along these directions are not in the same sense. Thus there is doubling of periodicity of the elementary pseudocubic lattice along all the three directions leading to appearance of superlattice reflections in the powder patterns. The x displacements however resemble the x displacements of the structure proposed by Noheda et al [2] assuming Cm space group. Their magnitudes are also almost identical within ± 0.01.

The refined cell parameters for the Cc space group can be converted into the equivalent cell parameters for the Cm space group using the transformation matrix given in eq. (1). These values are listed at the bottom of Table 2. It is evident that the two sets of cell parameters are also quite close to each other implying thereby that if one ignores the superlattice peaks in Fig.2, as was the case with Noheda et al [2] in their XRD studies, one would land up in the space group Cm.

## IV. CONCLUSIONS:

We have confirmed that the tetragonal phase of PZT with x=0.520 has a superlattice phase at low temperatures with space group Cc. This superlattice phase can result from the monoclinic Cm phase discovered by Noheda et al [2] by an antiferrodistortive phase transition. Further, this superlattice phase coexists with a secondary phase whose space group is also Cm but the equivalent perovskite cell parameters of this secondary phase bear pseudorhombohedral character [10] in contrast to the pseudotetragonal character of the equivalent perovskite cell parameters of the Cm phase discovered by Noheda et al [2]. The R3c space group proposed by Frantti et al [7] for the superlattice phase cannot account for the peak positions of the superlattice reflections and is therefore discarded. If one ignores the superlattice peaks, as was the case with Noheda et al [2] in their XRD studies, the most plausible space group of the low temperature phase becomes Cm.

**ACKNOWLEDGEMENT:** Ragini and Akhilesh acknowledge CSIR for the award of a Senior Research Fellowship and Research Associateship, respectively. Rajeev Ranjan is grateful to Department of Science and Technology, Govt. of India for financial support.

**Table 1.** Refined structural parameters and agreement factors for Pb(Zr$_{0.520}$Ti$_{0.480}$)O$_3$ at 10K using coexistence of two monoclinic phases with Cc and Cm space groups.

| Atoms | Space group : Cc | | | | Space group : Cm | | | |
|---|---|---|---|---|---|---|---|---|
|  | X | y | z | B(Å$^2$) | x | y | z | B(Å$^2$) |
| Pb | 0.00 | 0.25 | 0.00 | 0.84(6) | 0.0 | 0.0 | 0.0 | 1.6(1) |
| Zr/Ti | 0.215(1) | 0.24(1) | 0.692(4) | 0.1(2) | 0.52(1) | 0.0 | 0.44(1) | 0.0(3) |
| O1 | -0.045(1) | 0.238(4) | 0.418(2) | 0.3(1) | 0.577(4) | 0.00 | -0.078(7) | 0.5(7) |
| O2 | 0.187(2) | 0.486(3) | -0.078(4) | 1.6(4) | 0.290(7) | 0.253(2) | 0.415(6) | 0.6(4) |
| O3 | 0.196(1) | -0.013(2) | -0.103(3) | 0.0(1) |  |  |  |  |
| | a=10.004(7)Å, b=5.701(5)Å, c= 5.719(6)Å, β=124.36(5)°, R$_B$=5.22, | | | | a=5.758(1)Å, b=5.729(1)Å, c=4.0930(7)Å, β=90.59(1), R$_B$=4.19 | | | |
| % Molar | 65.18(1) | | | | 34.82(1) | | | |
| | R$_P$=7.21, R$_{W-P}$=9.96, R$_{exp}$=8.81 | | | | | | | |

**Table 2.** Comparison of the coordinates of the Cc phase after appropriate transformation to equivalent Cm phase type axes with those given by Noheda et al [2] for the Cm phase.

| | Our results | | | Noheda et al's results | | |
|---|---|---|---|---|---|---|
| Atoms | x | y | z | x | y | z |
| Pb | 0.00 | 0.00 | 0.00 | 0.00 | 0.00 | 0.00 |
| Zr/Ti | 0.523(5) | - 0.01(1) | 0.430(2) | 0.5230(6) | 0.00 | 0.4492(4) |
| O1 | 0.537(3) | - 0.012(4) | -0.090(2) | 0.5515(23) | 0.00 | - 0.0994(24) |
| O2 | 0.265(6) | 0.236(3) | 0.374(4) | 0.2880(18) | 0.2434(20) | 0.3729(17) |
| O3 | 0.299(4) | 0.737(2) | 0.392(2) | 0.2880(18) | 0.7566(20) | 0.3729(17) |
| a = 5.7190Å, b = 5.7021 Å, c = 4.129 Å, β =90.51$^0$ | | | | a=5.72204 Å, b=5.70957 Å, c=4.13651 Å, β =90.498$^0$ | | |

**Figure Captions**

1. Observed (open circles), calculated (solid lines) and difference (bottom of the figures) profiles of powder neutron diffraction pattern of Pb(Zr$_{0.520}$Ti$_{0.480}$) O$_3$ at 10K for the 311 and 511 superlattice reflections and 200 and 222 perovskite reflections. The indices for superlattice reflections are with respect to a doubled pseudocubic cell while the indices for other reflections are with respect to the elementary perovskite cell. The short vertical bars represent the peak positions of the various phases.

2. Observed (dots), calculated (solid lines) and difference (bottom of the figures) profiles of powder neutron diffraction pattern of Pb(Zr$_{0.520}$Ti$_{0.480}$) O$_3$ in the 2θ range 20 to 130 degrees refined with space group Cc+Cm at 10K. The short vertical bars represent the peak positions of the two phases.

3. Projection of atomic positions of the Cm and Ia ($\equiv$ Cc) phases down the [001] axis. Filled circles, open circles and dots represent Pb, O and Ti/Zr atoms, respectively. The vectors $\mathbf{a_m}$ and $\mathbf{b_m}$ represent the monoclinic axes and, $\mathbf{a_p}$ and $\mathbf{b_p}$ represent the pseudocubic axes in the layer. The small arrows show the directions only (and not the magnitudes) of the displacements of various atoms with respect to their symmetric perovskite positions. The y-displacements of O atoms along the pseudocubic [100], [010] and [001] are in opposite sense revealing doubling of the elementary pseudocubic subcell in all the three directions. The x displacements in both the phases are identical and do not lead to doubling of the pseudocubic subcell.

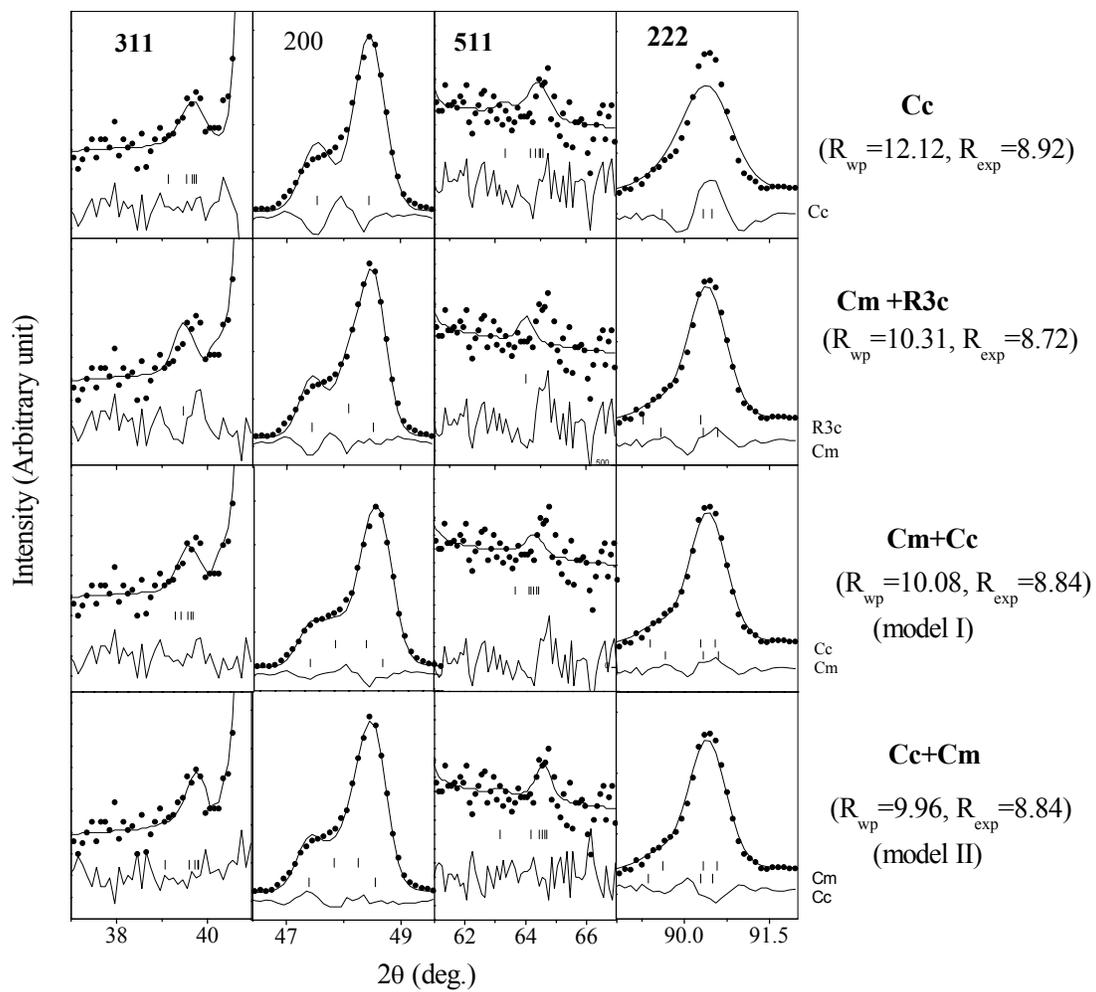

Fig. 1

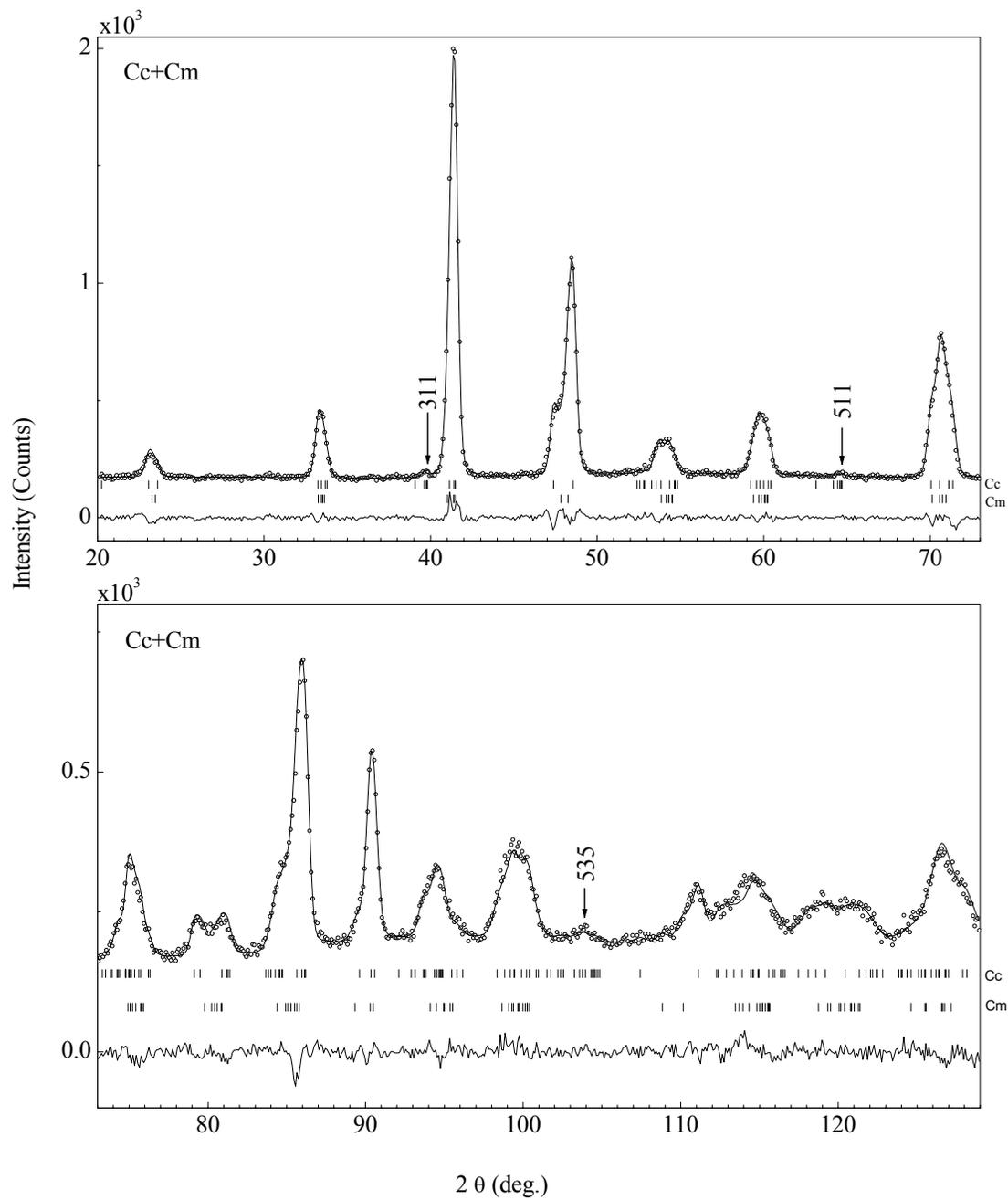

Fig. 2

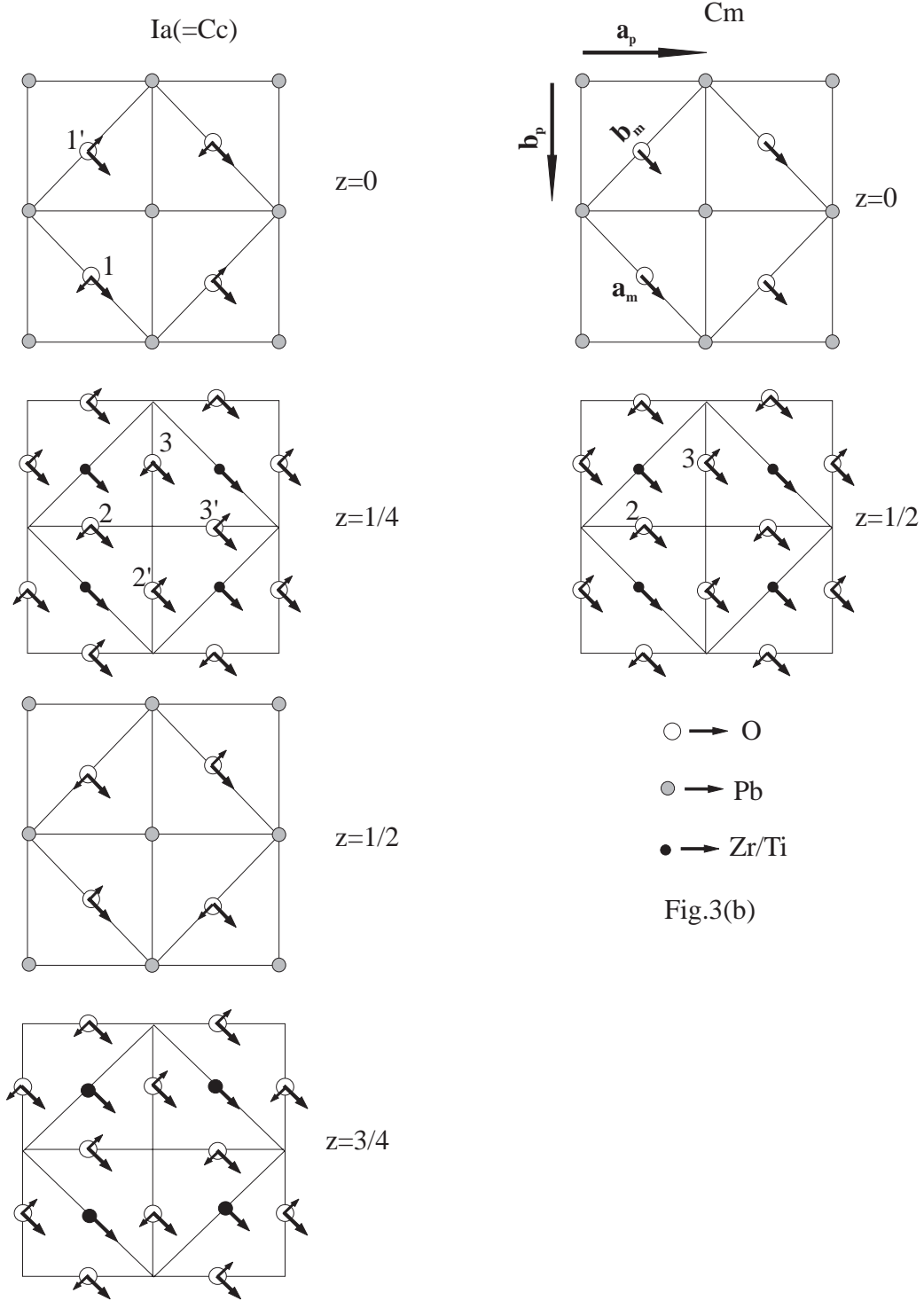

Fig.3(a)

Fig.3(b)